\documentclass[aps,prl,preprint,nofootinbib,superscriptaddress,balancelastpage,letterpaper]{revtex4}

\usepackage[T1]{fontenc} 
\usepackage[utf8]{inputenc}

\usepackage{graphicx}
\usepackage{graphics}
\usepackage[mathscr]{eucal}
\usepackage{amssymb} 
\usepackage{amsmath}
\usepackage{xspace}
\usepackage{listings}
\usepackage{ulem}
\usepackage{xcolor}
\usepackage{bm}         
\usepackage{array}
\usepackage{multirow}   
\usepackage{booktabs}   
\usepackage{mleftright}

\usepackage{bbold}
\usepackage{amsmath,amscd}
\usepackage{slashed}
\usepackage{placeins}
\usepackage{soul}
\usepackage{braket}
\usepackage{tikz-cd}
\usepackage{empheq}
\usepackage[most]{tcolorbox}
\usetikzlibrary{arrows, shapes,shadows}
\usepackage{mathtools,slashed}
\usepackage{dsfont}
\usepackage[makeroom]{cancel}
\usepackage{multirow}

\usepackage{amssymb}
\usepackage[capitalise]{cleveref}

\crefname{section}{Sec.}{Secs.}
\crefname{table}{Tab.}{Tabs.}
\crefname{figure}{Fig.}{Figs.}
\crefname{equation}{Eq.}{Eqs.}
\crefname{appendix}{Appendix\ }{Appendix\ }

\DeclareMathAlphabet{\mathpzc}{OT1}{pzc}{m}{it}

\begin{document}

\title{Scaling of high-energy elastic scattering and the observation of Odderon}

\author{T.~Cs\"org\H{o}}
\email{tcsorgo@cern.ch}
\affiliation{Wigner RCP, H-1525 Budapest 114, POBox 49, Hungary}
\affiliation{{
SZIE} KRC, H-3200 Gy\"ongy\"os, M\'atrai \'ut 36, Hungary}

\author{T.~Nov\'ak}
\email{novak.tamas@szie.hu}
\affiliation{{
SZIE} KRC, H-3200 Gy\"ongy\"os, M\'atrai \'ut 36, Hungary}
 
\author{R.~Pasechnik}
\email{Roman.Pasechnik@thep.lu.se}
\affiliation{Department of Astronomy and Theoretical Physics, Lund University, 
221 00 Lund, Sweden}

\author{A.~Ster}
\email{ster.andras@wigner.hu}
\affiliation{Wigner RCP, H-1525 Budapest 114, POBox 49, Hungary}

\author{I.~Szanyi}
\email{iszanyi@cern.ch}
\affiliation{Wigner RCP, H-1525 Budapest 114, POBox 49, Hungary}
\affiliation{E\"otv\"os University, H-1117 Budapest, P\'azm\'any P. s. 1/A, Hungary}

\begin{abstract}
We provide a statistically significant observation of the elusive Odderon exchange, based on 
novel and model-independent analysis of the scaling properties of the differential cross sections 
of elastic $pp$ and $p\bar p$ scattering in the TeV energy range. We report the statistical 
significance of the observed Odderon signal at the level of 6.26 $\sigma$.  
\end{abstract}

\maketitle

In 1973, Lukaszuk and Nicolescu~\cite{Lukaszuk:1973nt} proposed that a noticeable crossing-odd
contribution called Odderon may be present in the scattering amplitude of elastic proton-proton
($pp$) and proton-antiproton ($p\bar p$) scattering at asymptotically high energies. In the field 
theory of strong interactions, quantum chromodynamics (QCD), the Odderon exchange corresponds 
to the $t$-channel exchange of a color-neutral gluonic compound state consisting of an odd number 
of gluons, as elaborated by Bartels, Lipatov and Vacca in Ref.~\cite{Bartels:1999yt}. Although 
more than 20 years have passed since the theoretical prediction of the Odderon in QCD, and 
over 46 years since the Odderon concept has been introduced in Regge phenomenology, 
the Odderon remained elusive so far due to lack of a definitive experimental evidence.
A direct way to probe the Odderon in elastic scattering is by comparing the differential 
cross-section of particle-particle and particle-antiparticle scattering at the same and sufficiently high energy~\cite{Jenkovszky:2011hu,Ster:2015esa}.
The first search performed at the ISR energy of $\sqrt{s}=53$ GeV in 1985~\cite{Breakstone:1985pe} 
resulted in an indication of the Odderon at the 3.35 $\sigma$ significance level.
That analysis, however, did not utilize all the available data in the overlapping acceptance of 
the $pp$ and $p\bar p$ measurements.
Furthermore, at such a low energy the Reggeon exchanges are expected to 
play a significant role rendering the Odderon search at the ISR rather inconclusive.

Recently, the TOTEM Collaboration published a series of important papers investigating 
the properties of elastic $pp$ scattering in the LHC energy range between $\sqrt{s} = 2.76$ 
and $13$ TeV~\cite{Antchev:2017dia,Antchev:2017yns,Antchev:2018edk,Antchev:2018rec}. 
An increase of the total cross section, $\sigma_{\rm tot}(s)$, associated with 
a decrease of the real-to-imaginary ratio, $\rho(s)$, with energy, first identified 
at $\sqrt{s} = 13$ TeV \cite{Antchev:2017dia,Antchev:2017yns} indicated a possible 
Odderon effect triggering an intense debate~\cite{Khoze:2017swe,Samokhin:2017kde,Csorgo:2018uyp,Broilo:2018qqs,Pancheri:2018yhd,Goncalves:2018nsp,Selyugin:2018uob,Khoze:2018bus,Broilo:2018els,Troshin:2018ihb,Dremin:2018uwt,Martynov:2018nyb,Martynov:2018sga,Shabelski:2018jfq,Khoze:2018kna,Hagiwara:2020mqb,Contreras:2020lrh,Gotsman:2020mkd}. 
The persistent diffractive minimum-maximum structure in the $t$-dependent profile of $d\sigma/dt$ 
in elastic $pp$ collisions observed by the TOTEM at $\sqrt{s}$ = 2.76, 7 and 13 TeV, and 
the lack of such structure in elastic $p\bar p$ collisions measured by D0~\cite{Abazov:2012qb} 
at $\sqrt{s} = 1.96$ TeV, indicate a qualitatively clear Odderon effect \cite{Csorgo:2018uyp}.
Thus the TOTEM collaboration concluded in Ref.~\cite{Antchev:2018rec} as follows:
{\it ``Under the condition that the effects due to the energy difference between TOTEM and D0 can be 
neglected, the result provides evidence for a colourless 3-gluon bound state exchange in the $t$-channel 
of the $pp$ elastic scattering''}. However, no conclusive, quantitative experimental 
results were published so far with a statistically significant evidence for an Odderon
discovery. 

In this work, we present a definitive and statistically significant Odderon observation. This result 
is based on a re-analysis of already published D0~\cite{Abazov:2012qb} and TOTEM~\cite{Antchev:2013gaa,Antchev:2018edk,Antchev:2018rec} data sets, without the use of 
any fitting function or theoretical input. Namely, we compare pairwise the scaling functions 
constructed at different energies based upon the available data and look for statistically 
significant differences within any pair of TeV-scale $pp$ and $p\bar p$ data sets depending 
on the collision energy.
\begin{figure*}[hbt]
\begin{center}
\begin{minipage}{1.0\textwidth}
 \centerline{
 \includegraphics[width=0.48\textwidth]{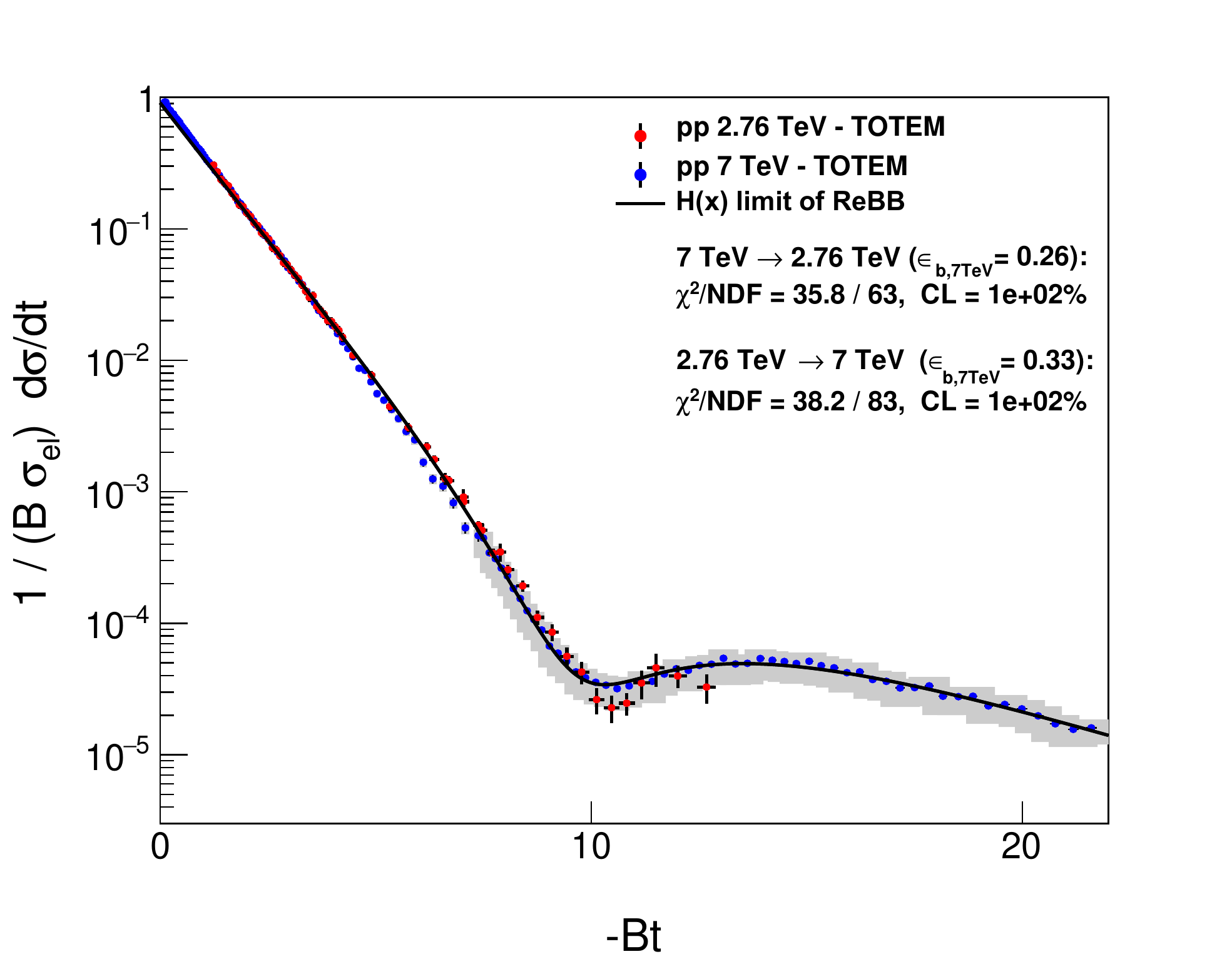}
 \includegraphics[width=0.48\textwidth]{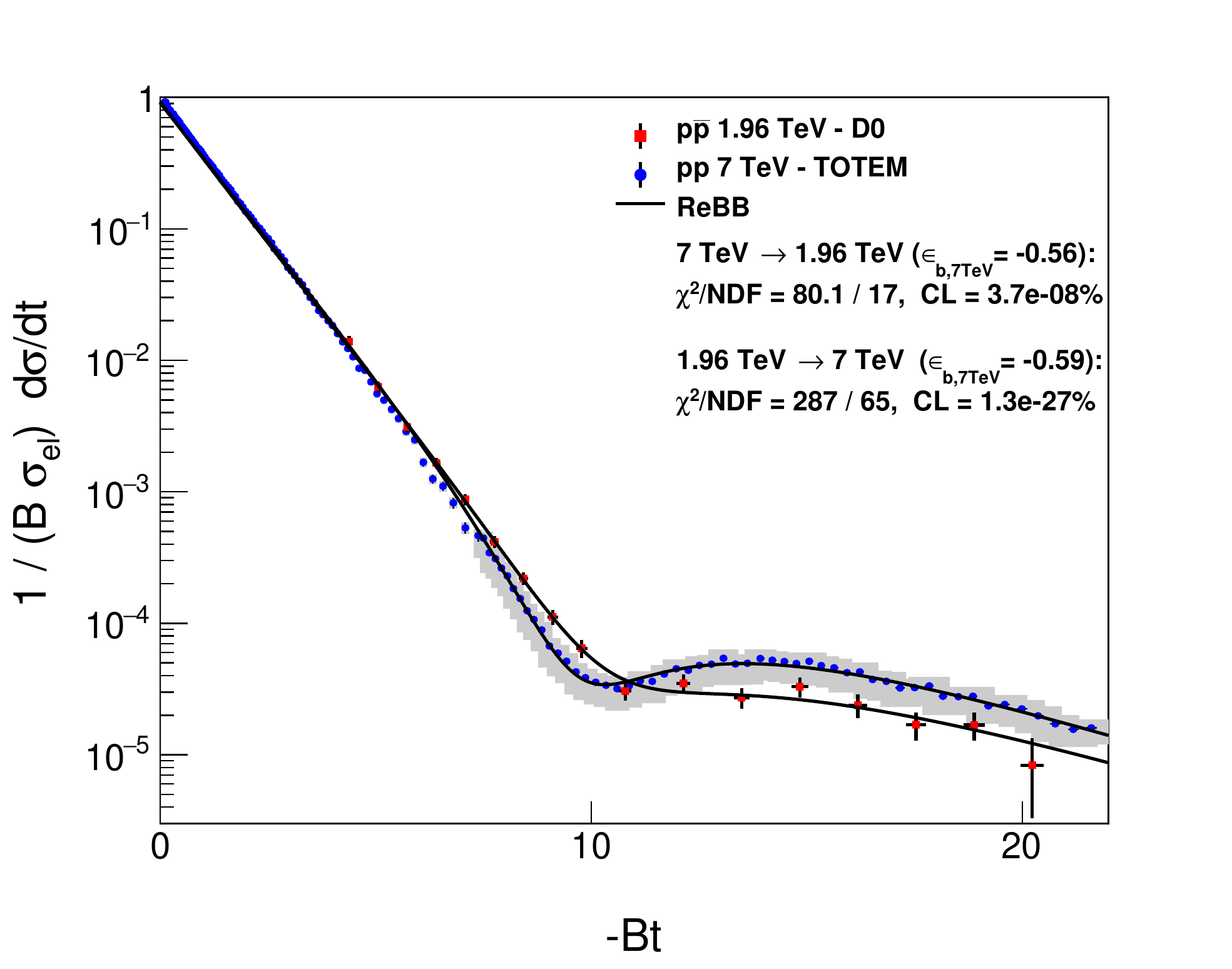}
 }
\vspace{-0.5truecm}
\end{minipage}    
\end{center}
\caption{ 
Left panel indicates that for $pp$ elastic scattering the $H(x)$ scaling function for $x = - t B$ 
is energy independent in the energy range of $\sqrt{s} = 2.76 - 7$ TeV. Vertical and horizontal 
lines on each point stand for the corresponding type A errors. Grey vertical bars represent 
the type B (vertical and horizontal) errors.  
The right panel indicates a statistically significant difference between the $H(x)$ scaling 
functions for elastic $pp$ collisions at $\sqrt{s} = 7$ TeV and that of $p\bar p$ collisions 
at $\sqrt{s} = 1.96$ TeV at the level of at least 6.26 $\sigma$. Here, $X\to Y$ denotes projections 
by exponential interpolation between the two adjacent data points of the data set $X$ to get its $H(x)$ 
at the same $x$ as that of the other data set $Y$ to be able to compare them via $\chi^2$-method 
as described in the text.
On the left panel, solid line indicates the $H(x)$ scaling limit of the ReBB model~\cite{Bialas:2006qf,Nemes:2015iia,Csorgo:2020wmw}.
On the right panel, solid line indicates the full ReBB model result, including scaling violations~\cite{Bialas:2006qf,Nemes:2015iia,Csorgo:2020wmw}. Both solid lines
correspond to Refs.~\cite{Nemes:2015iia,Csorgo:2020wmw}.
}
\label{fig:H(x)-Odderon}
\end{figure*}

Our analysis is based on a novel scaling {
law} of elastic $pp$ ($p\bar p$) 
scattering~\cite{Csorgo:2019ewn}. We utilize a new kind of scaling function, 
$H(x,s) = (1/B \sigma_{\rm el}) d\sigma/dt$, where $x = -t B$, and
\begin{equation}
     \sigma_{\rm el} = \int_{0}^\infty d|t| \,\, \frac{d\sigma}{dt} \,, \quad
     B = \frac{d}{dt} \ln \frac{d\sigma}{dt}\Big|_{t\to 0} \,, 
     \label{e:sigmael}
\end{equation}
that has been found to be energy independent, $H(x,s_1) = H(x,s_2)$, in elastic $pp$ collisions 
in a few TeV scattering domain, within the acceptance of TOTEM measurements at $2.76$, $7$ (and $8$) TeV
as demonstrated in the left panel of Fig.~\ref{fig:H(x)-Odderon}. This $H(x)$ scaling removes 
the trivial energy dependent terms, due to the known $s$-dependence of the elastic slope $B(s)$, 
the elastic and total cross-sections $\sigma_{\rm el}(s)$ and $\sigma_{\rm tot}(s)$, and 
the real-to-imaginary ratio $\rho(s)$~\cite{Csorgo:2019fbf}.
For $pp$ elastic scattering the $H(x,s)$ scaling function for $x = - t B$ 
is energy independent in the energy range of $\sqrt{s} = 2.76 - 7$ TeV, as shown on the left panel of Fig.
~\ref{fig:H(x)-Odderon}: within statistical errors,  the agreement 
corresponds to the confidence level (CL) of 99\%. Combining both the statistical and 
systematic errors, the agreement  corresponds to a $\chi^2/{\rm NDF} = 35.8/63$ in 
$1.2 < x < 12.7$, with a CL = 100 \%. A quantitative Odderon effect is then determined 
by a projection of the $pp$ and $p\bar p$ data and the corresponding uncertainties 
to the same $x(s)$  values at different energies. 
{
The $7\to 1.96$ TeV and $1.96\to 7$ TeV projections correspond to keeping the 
measured $x$ values at $\sqrt{s} = $ 1.96 and 7 TeV, respectively, and determining by interpolation
the $H(x,s)$ scaling functions at these $x$ values, but at the other energy, $\sqrt{s} = $ 7 and 1.96 TeV, respectively.}
The right panel of Fig.~\ref{fig:H(x)-Odderon} compares the $H(x)$ scaling function
of elastic $pp$ collisions at $\sqrt{s} = 7$ TeV to that of $p\bar p$ collisions 
at $\sqrt{s} = 1.96$ TeV. In this case, adopting the method of Ref.~\cite{Adare:2008cg}, 
the confidence level of the agreement of the $H(x)$ scaling functions  
is found to be maximum $3.7$ $\times$ $ 10^{-8}$\%, with a minimum of $\chi^2/{\rm NDF} = 80.1/17$. 
Hence, the difference between these scaling functions is statistically significant 
and represents our main result for the Odderon observation as at least 6.26 $\sigma$ effect,
with probability, $P = 1-{\rm CL} = 0.999999963$, in the $5<x\lessapprox 20$ acceptance. 
This is a conservative result as we find that this value is robust (can only increase) 
for the variation of the procedure and the $\chi^2$ definition. To guide the eye, 
solid lines, representing the ReBB model results~\cite{Bialas:2006qf,Nemes:2015iia,Csorgo:2020wmw} 
are also added to both panels of Fig.~\ref{fig:H(x)-Odderon}.

As a cross-check, we have tested the validity of the $H(x)$ scaling versus the TOTEM  
data on elastic $pp$ scattering at $\sqrt{s_1} = 7$  TeV and TOTEM preliminary data and errors at $\sqrt{s_2} = 8$
TeV~\cite{Kaspar:2018ISMD}. We find  that $H(x,s_1) = H(x,s_2)$ at CL  $\approx 100 \%$ in the $5 \leq x \leq 20$-range.
Due to a lack of direct measurements of $pp$ and $p\bar p$ collisions at exactly the same energy in 
the TeV region, we utilize the energy independence of the $H(x)$ scaling in the LHC energy 
range of $1 \lesssim \sqrt{s} \lesssim 8$ TeV, to evaluate the characteristics of the elastic 
$pp$ scattering at the D0 energy of $\sqrt{s} = 1.96$ TeV.

We have also cross-checked this scaling behaviour at ISR energies and found that all the differential 
cross sections of elastic $pp$ scattering, measured at the ISR energy range of $\sqrt{s} = 23.5$ -- $62.5$ 
GeV \cite{Amaldi:1979kd,Breakstone:1984te}, can approximately be scaled to the same universal curve~\cite{Csorgo:2019ewn}.
We have also studied the $H(x)$ scaling 
for elastic $p\bar p$ collisions in the energy range of $\sqrt{s} = 0.546$ -- $1.96$ TeV and 
found that in this case, the scaling is limited to the diffractive cone, $x \le 10$ only, where 
$H(x)$ $\approx$ $\exp(-x)$, but in $p\bar p$ collisions the $H(x)$ scaling is strongly 
and qualitatively violated for $x > 10$ values. However, the valid $H(x)$ scaling in $pp$ scattering allowed us to scale down the higher energy TOTEM $pp$ data from $\sqrt{s} = 7$ TeV to $\sqrt{s} = 1.96$ TeV and directly compare it to the D0 $p\bar p$ data.

The quantification of the Odderon significance is based on a method developed by the PHENIX 
collaboration in Ref.~\cite{Adare:2008cg} using a specific $\chi^2$ definition that effectively 
diagonalizes the covariance matrix. In the PHENIX formulation, the experimental 
data are compared to a theoretical calculation. In our analysis, we adapt the PHENIX 
method for comparison of one set of data directly to another set of data, without using any theory or fitting functions. 
Following the PHENIX method, we classify the experimental errors of a given data set into three different types: (i) type A, 
point-to-point fluctuating (uncorrelated) systematic and statistical errors, (ii) type B errors 
that are point-to-point dependent, but 100\% correlated systematic errors, and (iii) type C errors, 
that are point-to-point independent, but fully correlated systematic errors~\cite{Adare:2008cg}
to evaluate the significance of correlated data, when the covariance matrix is not publicly available. Since the $t$-dependent systematic errors in TOTEM measurements are almost 100 \% correlated, 
we classified them as type B errors, while the $t$-independent overall normalization errors are type C errors, 
and the statistical errors are type A errors.

The source of the TOTEM $pp$ differential cross section data, measured at $\sqrt{s} = 7$ TeV, is Ref.~\cite{Antchev:2013gaa}. In addition, the values of $|t|$ were determined together with their errors of type A and B  
as given in Table 5 of Ref.~\cite{Antchev:2013haa} and Table 3 of Ref.~\cite{Antchev:2011zz}. The $t$-independent, type-C errors cancel from the $H(x)$ scaling functions, as they multiply both the numerator and the denominator of $H(x)$. At $\sqrt{s} = 2.76$ TeV, in Ref.~\cite{Antchev:2018rec}, the TOTEM Collaboration published the $pp$ differential cross section data with separated type-A and type-B errors. However, the D0 collaboration did not publish type-B errors for its differential cross-section data  at $\sqrt{s} = 1.96 $ TeV~\cite{Abazov:2012qb}. We have thus fixed the correlation coefficient of these D0 type-B errors to zero. 
{
The input values of the nuclear slope parameters $B$ and the elastic cross sections $\sigma_{\rm el}$ are summarized in Table~\ref{table:B-sigma},  together with the appropriate references.}

We define the significance of the agreement between the data set $D_1$ and the projection $D_{21} = D_2 \to D_1$ of {
data set $D_2$ to $D_1$}  in their overlapping acceptance, 
with the the following $\chi^2$ definition \cite{Csorgo:2019ewn}:
\begin{eqnarray*}
\chi^2_{2 \rightarrow 1} & = & \sum_{j=1}^{n_{21}}
    \frac{ (d_1^j +\epsilon_{b,1} e_{B,1}^j - 
    d_{21}^j - \epsilon_{b,21} e_{B,21}^j)^2 }
    {({\tilde e}_{A,1}^j)^2 + ({\tilde e}_{A,21}^j)^2} +
    \epsilon_{b,1}^2 + \epsilon_{b,21}^2 \,, 
    \\
\tilde{e}_{A,k}^j  & = &  e_{A,k}^j \frac{d_k^j + 
\epsilon_{b,k} e_{B,k}^j}{d_k^j} \,, \\
e_{M,k}^j  & = & \sqrt{(\sigma_{M,k}^j)^2 + (d^{\prime,j}_k)^2 (\delta_{M,k}^j x)^2}  \,,
\end{eqnarray*}
where $n_{21}$ is the number of data points $d_{21}^j$ in $D_{21}$ indexed by $j$, 
the same as in $D_1$ but remaining in the overlapping acceptance of $D_{1,2}$ sets, 
$e_{M,k}^j$, $k=1,21$, are the type $M = A,B$ errors found in terms of 
the type-M vertical errors on data point $j$, 
$\sigma_{M,k}^j$, added in quadrature with the corresponding type-M vertical errors that were evaluated from 
the corresponding errors on the horizontal axis $x$ with the scaled variance method, $d^{\prime,j}_k \delta_{M,k}^j x$, 
where $d^{\prime,j}_k$ 
stands for the numerical derivative of the measured quantity in data set $D_k$ at the 
point $j$ in the common acceptance and $\delta_{M,k}^j x$ is the $j$-dependent type-M horizontal error.
The overall correlation coefficients of the type B errors $e_{B,k}^j$
of $D_k$ data sets are  
denoted by $\epsilon_{b,k}$.
\begin{table*}[htb]
\begin{center}
\begin{tabular}{l l l}
$\sqrt{s}$ (GeV)  &\,\,\,\, $\sigma_{\rm el}$ (mb)  &\,\,\, $B$ (GeV$^{-2}$) \\
\hline
 1960 ($p\bar{p}$)  &\,\,\,\, 20.2  $\pm$ $1.7^{A} $ $\pm$ $14.4\%^{C}$ [*]  
 & \,\,\, 16.86 $\pm$ $0.1^{A}$  $\pm$ $0.2^{A}$  ~\cite{Abazov:2012qb} \\
 2760 ($pp$)          &\,\,\,\, 21.8  $\pm$ $1.4^{A} $  $\pm$ $6.0\%^{C}$  ~\cite{Nemes:2017gut,Antchev:2018rec}
                     & \,\,\, 17.1  $\pm$ $0.3^{A}$ ~\cite{Antchev:2018rec}     \\
 7000 ($pp$)          &\,\,\,\, 25.43 $\pm$ $0.03^{A}$ $\pm$ $0.1^{B}$ $\pm$ $0.31^{C}$ $\pm$ $1.02^{C}$ ~\cite{Antchev:2013gaa}  
 & \,\,\, 19.89 $\pm$ $0.03^{A}$ $\pm$ $0.27^{B}$ ~\cite{Antchev:2013gaa} \\
\hline
\end{tabular}
\end{center}
\caption {Summary table of the elastic cross-sections $\sigma_{\rm el}$ and the nuclear slope parameters $B$, 
with references. We have indexed with superscripts $A,B,C$ the type A,B,C errors, respectively. The value and the type A error of the elastic cross-section $\sigma_{\rm el}$ at $\sqrt{s} = 1.96$ TeV [*] 
is obtained from a low $-t$ exponential fit to the data of Ref.~\cite{Abazov:2012qb}, while the type C error 
is from Ref.~\cite{Abazov:2012qb}. The statistical and systematic errors of $d\sigma/dt$ 
data at $\sqrt{s} = 1.96$ TeV were  added in quadrature in Ref.~\cite{Abazov:2012qb}, therefore 
it was done in case of the elastic slope $B$ as well, providing a combined type A error 
$\delta^{A} B = 0.224$ GeV$^{-2}$. At $\sqrt{s} = 2.76 $ TeV, Ref.~\cite{Antchev:2018rec} provides the total error on $B$, 
without decomposing it into type A and type B parts. Similarly, the error 
on the TOTEM preliminary value of the elastic cross section 
at $\sqrt{s} = 2.76$ TeV was not decomposed to type A and B errors in Ref.~\cite{Nemes:2017gut}, either. Hence, we treat 
these  as errors of type A:  this assumption yields a conservative estimate 
of the Odderon significance in our calculations.
}
\label{table:B-sigma}
\end{table*}

\begin{figure*}[htb]
\begin{center}
\begin{minipage}{1.0\textwidth}
 \centerline{
 \includegraphics[width=0.48\textwidth]{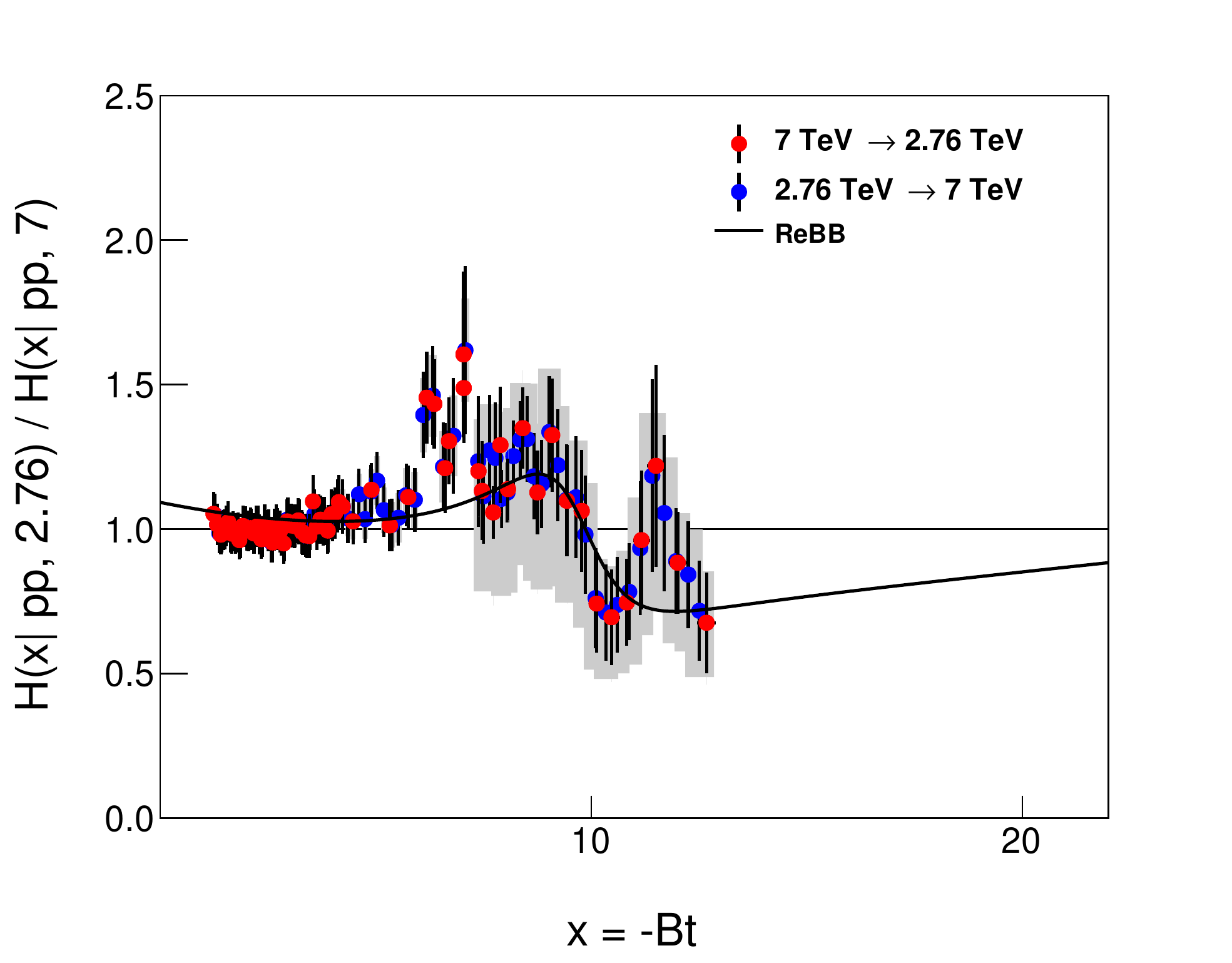}
 \includegraphics[width=0.48\textwidth]{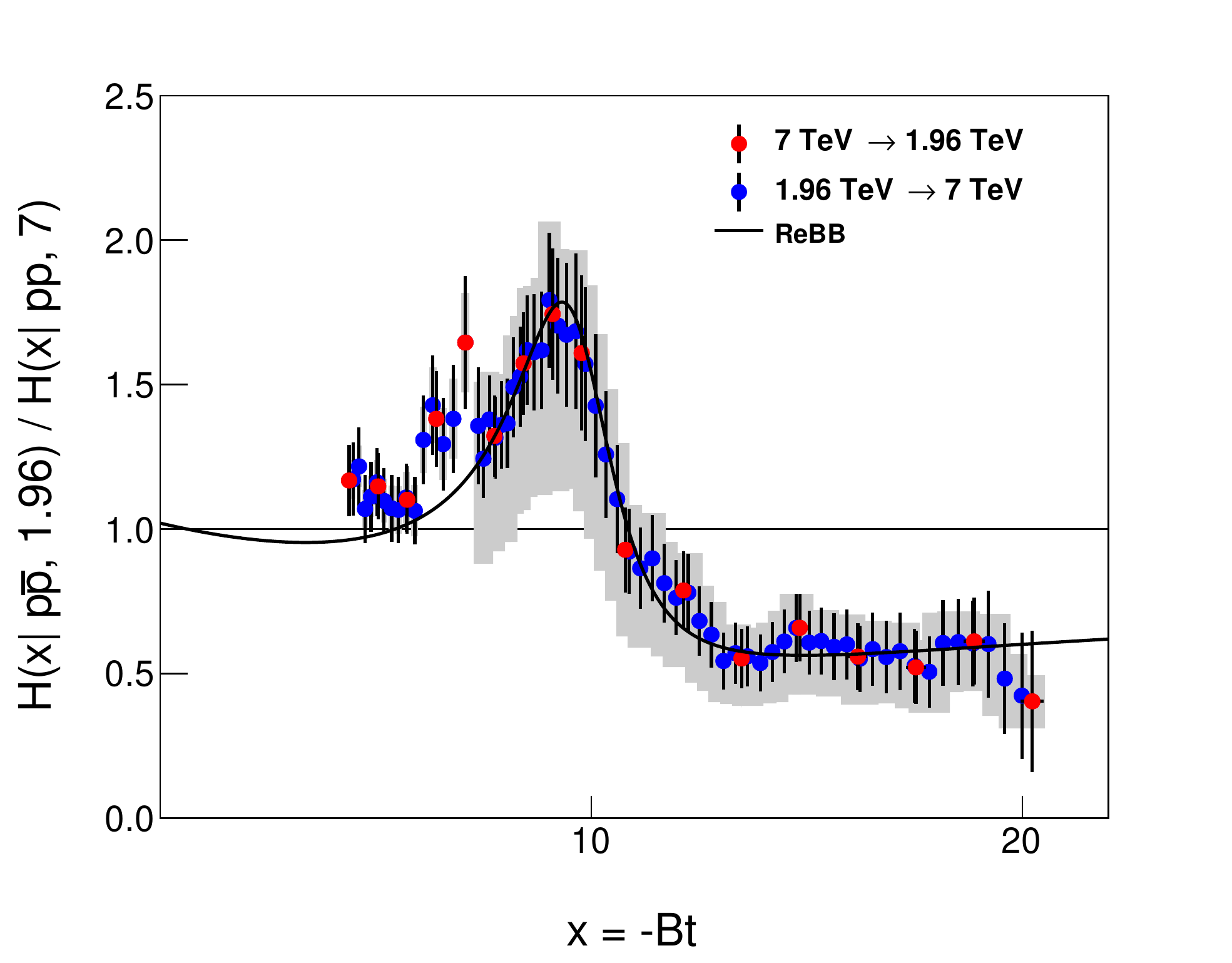}}
 \vspace{-0.5truecm}
\end{minipage}    
\end{center}
\caption{
Left panel indicates that for $pp$ elastic scattering the ratio of the scaling functions $H(x,s_1)/H(x,s_2)$, 
where $x = - t B$, $\sqrt{s_1}=2.76$ TeV and $\sqrt{s_2}=7$ TeV, is not inconsistent with unity within 
statistical errors, due to the energy independence of the $H(x)$ scaling in the 
$2.76 \le \sqrt{s_{1,2}} \le 7$ TeV energy range. 
The right panel indicates a statistically significant deviation from unity of the ratios of the $H(x)$ scaling functions 
for elastic $p\bar p$ collisions at $\sqrt{s} = 1.96$ TeV, and that of $pp$ collisions at $\sqrt{s} = 7$ TeV.
Notation and the experimental data are the same as in Fig.~\ref{fig:H(x)-Odderon}, but represented here as ratios.
On both panels, curved solid lines indicate the results of Refs.~\cite{Bialas:2006qf,Nemes:2015iia,Csorgo:2020wmw} . 
The straight solid black lines at unity correspond to the scaling limit in both panels.
}
\label{fig:H(x)-ratio-Odderon}
\end{figure*}

Let us also demonstrate the presence of the Odderon in elastic scattering in the TeV energy range in a new, qualitative manner by
representing the Odderon effect as a peak in deviation from {
the baseline of a background, normalized to unity. 
On both panels of 
Fig.~\ref{fig:H(x)-ratio-Odderon} we plotted, on a linear scale, the ratio of two differential cross-sections 
that decrease over five orders of magnitude. This way of plotting follows the good practices of the field, established 
e.g.~in Ref.~\cite{Breakstone:1985pe}. This plot magnifies the possible differences and any trends in the deviations 
in a much more transparent way as compared to Fig.~\ref{fig:H(x)-Odderon}.}
In Fig.~\ref{fig:H(x)-ratio-Odderon} (left) we demonstrate that the ratio of the $H(x,s)$ scaling 
functions for elastic $pp$ scattering at two distinct energies of $\sqrt{s} = 2.76$ and $7$ TeV is indeed {
not in}consistent 
with unity within statistical errors. This means that $H(x,s)$ with $x = - t B$ is energy-independent at least 
in the range of $2.76 \le \sqrt{s} \le 7$ TeV at CL = 99 \%. This also indicates that in the energy range of a few TeV, 
the trivial energy dependence is indeed scaled out from 
the differential cross section of elastic $pp$ scattering in the $H(x,s)\simeq H(x)=(1/B \sigma_{\rm el})d\sigma/dt$ function.
As a cross-check of uncertainties, we have considered two distinct directions of projection: direct $2.76\to 7$ TeV and 
inverse $7\to 2.76$ TeV denoted by blue and red central points, respectively, and no significant difference has been observed.
This shows remarkable stability of our results with respect to the details of the projection procedure.
We have also observed the same picture for the ratio of the $H(x)$ scaling functions at $\sqrt{s_1} = 7$ 
and $\sqrt{s_2}=8$ TeV, $H(x,s_1)/H(x,s_2)$, for $pp$ collisions.
A solid line is also added, to guide the eye and to
indicate the magnitude of the scaling violations in this $pp$ to $pp$ comparison, estimated with the help of 
Refs.~\cite{Bialas:2006qf,Nemes:2015iia,Csorgo:2020wmw}.

In Fig.~\ref{fig:H(x)-ratio-Odderon} (right) we present visible and statistically significant deviation from unity in the ratio of the
scaling functions  of $pp$ and $p\bar p$ elastic scattering.
The ratio of the $H(x)$ scaling functions is shown for elastic $p\bar p$ 
collisions at $\sqrt{s} = 1.96$ TeV over that of $pp$ collisions at $\sqrt{s} = 7$ TeV. As a cross-check, we show 
the results of two different projection procedures: direct $1.96\to 7$ TeV and inverse $7\to 1.96$ TeV denoted by blue 
and red central points, respectively. No significant variation with respect to the direction of projection has been found. 
In both ways, we observe a statistically significant Odderon effect as a peak in the $5<x<10$ region, followed by a factor 
of two suppression or decrease from unity in a broad range of $10\lessapprox x=-tB\lessapprox 20$. The statistical significance 
of the observed difference between the $pp$ and $p\bar p$ scaling functions has been found to be at least 6.26 $\sigma$, 
consistently with the result of a direct comparison of the scaling functions as shown in Fig.~\ref{fig:H(x)-Odderon}. 
A solid line is added to this panel too in order to estimate the magnitude of the $H(x)$ scaling violations in this 
$pp$ to $p\bar p$ comparison~\cite{Nemes:2015iia,Csorgo:2020wmw}.

{
One may wonder if this statistically significant and model independent Odderon signal is due to an extrapolation procedure from 
$\sqrt{s} = 7 $ TeV down to 1.96 TeV, that assumes the validity of the $H(x)$ scaling. 
Figs.~16 and 17 
in Ref.~\cite{Csorgo:2020wmw} address this question without any reference to the validity of the $H(x)$ scaling, but relying 
on the ReBB model of Ref.~\cite{Nemes:2015iia}. As these results include the terms that violate the $H(x)$ scaling, they allow 
not only for the evaluation of $pp$ $d\sigma/dt$ down to $\sqrt{s} = 1.96$ TeV, but also for the extrapolation of $p\bar p$ 
$d\sigma/dt$ up to $\sqrt{s} = 2.76$ TeV as well as to higher energies. These validated extrapolations result in a $\chi^2/\rm{NDF} 
= 24.28/13 $ and $100.35/20$ at $\sqrt{s} = 1.96$ and $2.76$ TeV, respectively. Within the ReBB model, the combined Odderon 
significance of the 1.96 $p\bar p$ and 2.76 TeV $pp$ data can be calculated from the combined $\chi^2/\rm{NDF} = 124.63/33$. 
This corresponds to a $\rm{CL} = 1.44 \times 10^{-10}$\%, and to a model-dependent Odderon significance of 7.08 $\sigma$. The same model
~\cite{Csorgo:2020wmw} implies that the domain of validity of the $H(x,s) = H(x,s_0)$ scaling includes the  $1.8 \leq \sqrt{s} \leq 8 $ TeV energy range, 
where both $\rho(s)$ ~\cite{Antchev:2017yns} and  $\sigma_{el}(s)/\sigma_{tot}(s)$~\cite{Csorgo:2019fbf} are, within errors, independent of $s$. For some observables, this $s$-range may extend down to $\sqrt{s} = 0.4$ TeV~\cite{Csorgo:2020wmw}. Our results can thus be cross-checked in elastic $pp$ collisions at $\sqrt{s} = 510$ GeV,  with the STAR detector at the RHIC accelerator~\cite{Adam:2020ozo}.
}

Our final, combined, conservative significance is an at least 6.26 $\sigma$, crossing-odd effect in the scaling properties of $pp$ and 
$p\bar p$ scattering, obtained without any reference to modeling and without removing (or adding) any of the published D0 or TOTEM
data points. If the high energy limit of proton is a black disk of gluons, elastic scattering becomes flavor-blind and the leading 
squared logarithmic energy-dependent terms become the same for all reactions, from $p\pi$ to $p\gamma$ 
scattering~\cite{Block:2011vz}. Our result of a statistically significant Odderon observation implies, 
that elastic scattering is not flavor blind even at the TeV scale. Instead of the traditional black 
disc picture~\cite{Block:2011vz}, our results thus support the newly emerging black ring picture 
of protons at asymptotically large energies~\cite{Csorgo:2019egs}.

{\it Acknowledgments:} 
We acknowledge inspiring and useful discussions with W. Guryn, G. Gustafson, V. A. Khoze, E. Levin, L. Lönnblad, M. Strikman, 
M. Sumbera  and  members of the D0 and TOTEM collaborations.
R.P. is 
supported by the Swedish Research Council grants No. 621-2013-4287 and 2016-05996, by the European 
Research Council (ERC) under the European Union's Horizon 2020 research and innovation programme (grant agreement No 668679), 
as well as by the Ministry of Education, Youth and Sports of the Czech Republic project LTT17018.
T. Cs., T. N., A. S. and I. Sz. were partially supported by the NKIFH grants No. FK-123842, FK-123959 and K-133046 as well as 
by the EFOP 3.6.1-16-2016-00001 grant (Hungary). 
Our collaboration has been supported by the COST Action CA15213 
(THOR).

\vfill\eject
\bibliography{Odderon-Letter}

\end{document}